# Growth and Characterization of Fe$_{0.95}$Se$_{0.6}$Te$_{0.4}$ Single Crystal

Anmol Shukla, Vishal K. Maurya and Rajendra S. Dhaka[a]

*Novel Materials and Interface Physics Laboratory,*
*Department of Physics, Indian Institute of Technology Delhi, Hauz Khas, New Delhi-110016 INDIA*

[a]Corresponding author: rsdhaka@physics.iitd.ac.in

**Abstract.** In this paper we present the single crystal growth of Fe$_{0.95}$Se$_{0.6}$Te$_{0.4}$ high T$_C$ superconducting sample by the modified Bridgman technique. The x-ray diffraction pattern shows the single crystal nature of the sample, as only *(00l)* peaks are detectable. The stoichiometric composition has been verified by energy dispersive x-ray analysis. The superconducting transition temperature at 14 K was confirmed through DC magnetization (ZFC-FC) and resistivity measurements. By analyzing the isothermal M-H curves, we determined the value of $H_{c1}(0) \sim 360\, Oe$ by extrapolating the data. The temperature coefficient of resistivity obtained using the power law fitting was found to be 0.6. The obtained Raman spectra at room temperature can be interpreted with the tetragonal crystal structure and space group P4/nmm.

## INTRODUCTION

The discovery of superconductivity (SC) in oxypnictides La(O$_{1-x}$F$_x$)FeAs at ~8K [1] has lead to vast interest in the realm of condensed matter physics. In earlier days of superconductivity research, it was highly prohibited to convolute SC with magnetic elements, such as Fe, Mn, and Cr. However, there are a number of unconventional superconductors exist in which superconducting state is found in the neighborhood of magnetic state. The trick is to suppress the magnetic state by pressure/doping and working on this line of thought Fe based superconductors were discovered and very soon superconductivity was identified in iron chalcogenides (namely, α-FeSe at 8 K [2]). All the Fe based superconductors share a common layered structure with planar layer of Fe atoms joined by pnictides or chalcogenides arranged in stacked sequence [2]. Similar to cuprate superconductors, the high T$_C$ superconductivity in these materials is thought to instigate from Fe-layered structure. However, an accurate theoretical description of high T$_C$ and the deviation from usual BCS theory are still under debate.

In a recent study, using angle resolved photoelectron spectroscopy; it has been shown that there is a presence of Dirac cone type helical surface states in FeSe$_{0.5}$Te$_{0.5}$ single crystals [3]. Due to these surface states a 3D topological superconductor (TS) surface can be capable of hosting Majorana fermions. The possibility of these special surface states in Fe based superconductors further invokes the search for high quality FeSe$_{0.5}$Te$_{0.5}$ single crystals with improved T$_C$. The relatively high T$_C$ (14 K at ambient pressure [4] and around 26.2 K under 2 GPa pressure [5]), and easy crystal synthesis process makes FeSe$_{1-x}$Te$_x$ best candidate to utilize this system for TS based applications. In this paper, we report the growth of high quality single crystal of a Se rich composition [6, 7], Fe$_{0.95}$Se$_{0.6}$Te$_{0.4}$ in Fe-Se series, by the modified Bridgman technique and its superconducting and optical properties.

## EXPERIMENTS

High quality single crystals of Fe$_{0.95}$Se$_{0.60}$Te$_{0.40}$ were synthesized by the modified Bridgeman method. High purity Fe pieces (99.99 wt. *%*), Te shots (99.999 wt. *%*), and Se shots (99.999 wt. *%*) were vacuum sealed into a quartz ampoule. Since, quartz ampoule often breaks during cooling process, double sealing was done. The double sealed quartz tube was vertically hanged inside a modified Bridgman furnace and heated up to 950°C at the rate of 50°C/hr, and held at 950°C for 36 hr, cooled down to 400°C at the rate of 6°C/hr and further held at 400°C for 18 h. This was followed by quenching the sample in cold water. We got shiny crystals, which were easily cleavable. A few millimetres sized rectangular crystals were cut with help of scalpel blade and preserved in vacuum desiccator for the measurements. The x-ray diffraction pattern of as grown single crystals was taken using *PANanalytical ARIES* diffractometer with Cu *Kα* radiation at room temperature. The SEM and EDX spectra were obtained by *RONTEC* model *QuanTax200* at room temperature. The low field DC magnetization measurements were performed using superconducting quantum interference device (SQUID) magnetometer. The resistivity measurements were performed in four probe more using a physical property measurement system (PPMS). The Raman spectrum was obtained using *Reinshaw inVia* conformal Raman microscope with an unpolarized laser (10 mW) of 532 and 325 nm excitation wavelengths and grating element of 2400 and 1800 line/mm, respectively.

## RESULTS AND DISCUSSION

The shiny and pristine crystals were cleaved [a photograph is shown in Fig. 1(a)]. Figure 1(b) shows the x-ray diffraction (XRD) pattern of as grown single crystals at room temperature, which shows that only (0 0 *l*) planes are present, confirming the growth has occurred along the *ab* plane and indicating good quality of grown crystals. The SEM image reveals layered morphology and confirms the layered growth of single crystals [shown in Fig. 1(c)] [8, 9]. The energy dispersive x-ray (EDX) analysis [Fig. 1(d)] found the composition $Fe_{0.95}Se_{0.60}Te_{0.40}$, which is very close to the starting nominal concentration.

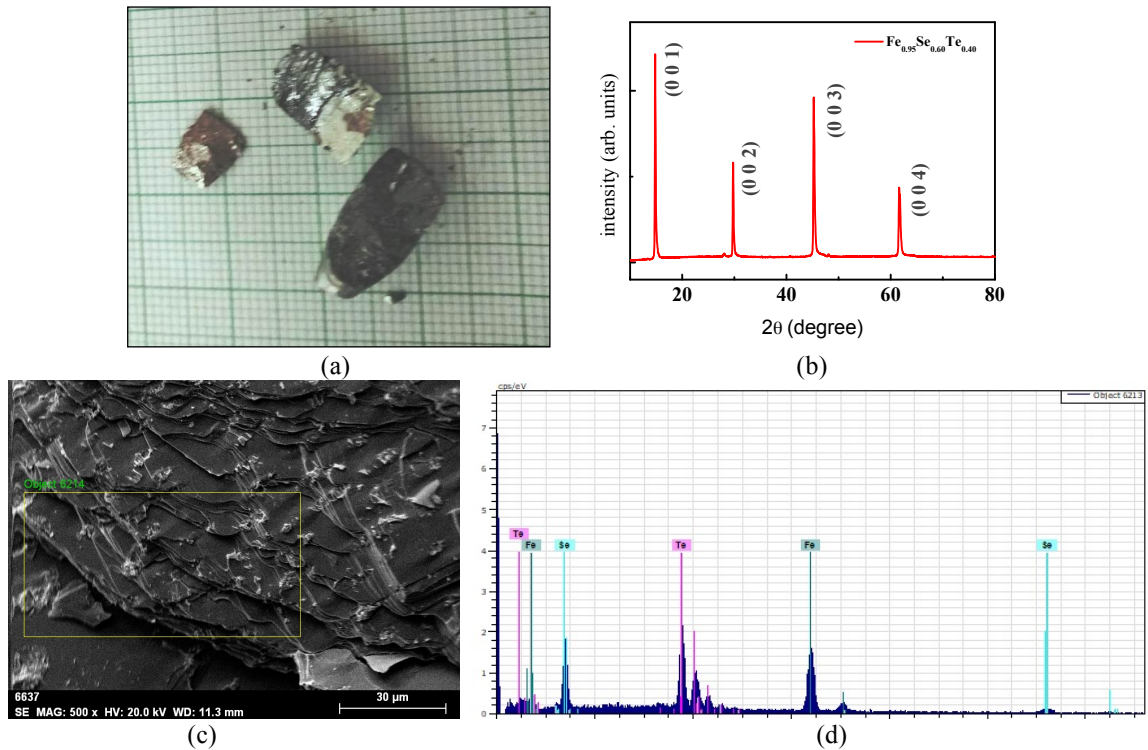

**FIGURE 1**: (a) A photograph of as grown $Fe_{0.95}Se_{0.6}Te_{0.4}$ single crystals on graph paper, each small box denotes area of 1 mm$^2$, (b) XRD pattern for $Fe_{0.95}Se_{0.6}Te_{0.4}$ crystals (x-rays incident upon *a-b* plane) at room temperature, (c) SEM image of crystals at 500X magnification, (d) EDX quantitative image of as grown single crystal.

Figure 2(a) shows the temperature dependent ZFC and FC dc-susceptibility curves measured from 4-20 K during heating cycle. The ZFC-FC data are taken at 10 Oe magnetic field, far below than reported $H_{c1}$ [8, 9], which make sure that the sample is in pure diamagnetic state and not in the mixed state. The superconducting $T_C$ = 14 K has been determined at the point where ZFC curve shows sudden drop in the magnetization. The isothermal magnetization hysteresis loop measured at 2 K is shown in Figure 2(b). This plot shows a clear evidence of type-II superconductivity. Further, broad opening of M-H curve till 1 Tesla suggests high critical current and high upper critical field. The evaluation of lower critical field, i.e. $H_{c1}$ of $Fe_{0.95}Se_{0.60}Te_{0.40}$ at 2 K is observed to be ~352 Oe by taking the deviation from the linearity criteria.

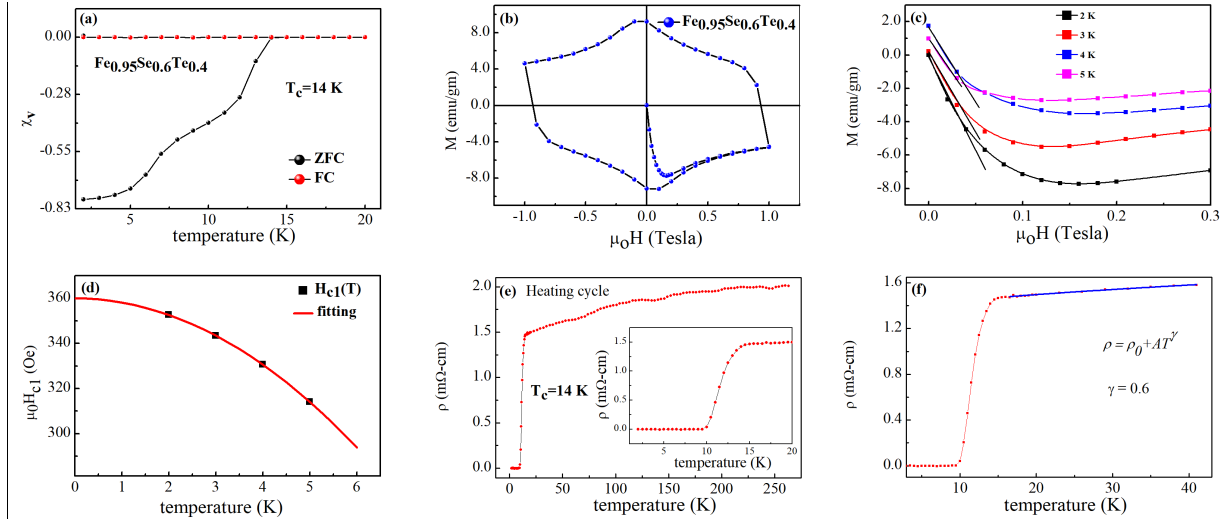

**FIGURE 2**: (a) ZFC and FC data during heat cycle shows superconducting $T_C$ at 14K, (b) isothermal curve at 2K shows a typical superconductor M-H loop, (c) M-H curves for increasing magnetic field values for the determination of $H_{c1}$, (d) fitting of $H_{c1}(T)$ data and extrapolation according to $H_{c1}(T) = H_{c1}(0)[1 - (T/T_C)^2]$ relation, (e) Temperature dependence of resistivity shows a superconducting transition around 14 K, and (f) the power law fitting above $T_C$ shown by blue line.

The lower critical fields at different temperatures were determined by analyzing the isothermal magnetization curves taken from 2 to 5 K [Fig. 2(c)]. The value of $H_{C1}(0)$ was calculated by extrapolating the data, using the single band relation; $H_{c1}(T) = H_{c1}(0)[1 - (T/T_c)^2]$ and shown in Fig. 2(d). The value of lower critical field; $H_{C1}(0)$ is estimated to be 360 Oe. In order to gain insight about the surface barrier, we have calculated first penetration field ($H_{c1}^*$). An easy relation between ($H_{c1}^*$) and ($H_{c1}$) given by, ($H_{c1}^*$) = ($H_{c1}$)/$\tanh\sqrt{0.36b/a}$ exists [9] for rectangular sample with, '*a*' and '*b*' as width and thickness of the sample, respectively. Using this formula, we found ($H_{c1}^*$) = 1123 Oe, where, '*a*'= 3.2 mm and '*b*'= 0.98 mm. The resistivity vs. temperature, shown in Fig. 2(e) displayed the onset of superconducting transition around 14 K. The resistivity above $T_c$ was fitted in the low temperature regime, up to 40 K, by the power law [Fig. 2(f)] $\rho = \rho_0 + AT^\gamma$. The temperature exponent of resistivity ($\gamma$) was found to be ~0.6. The value of exponent shows a marked deviation from the Fermi liquid behaviour ($\gamma = 2$). This non-Fermi liquid behaviour is similar to what is seen in the normal state of cuprate and other compositions in Fe-Se-Te superconductors [10]. The temperature independent part of the resistivity, namely, the residual resistivity ($\rho_0$) was found to be 1.33 mΩ-cm.

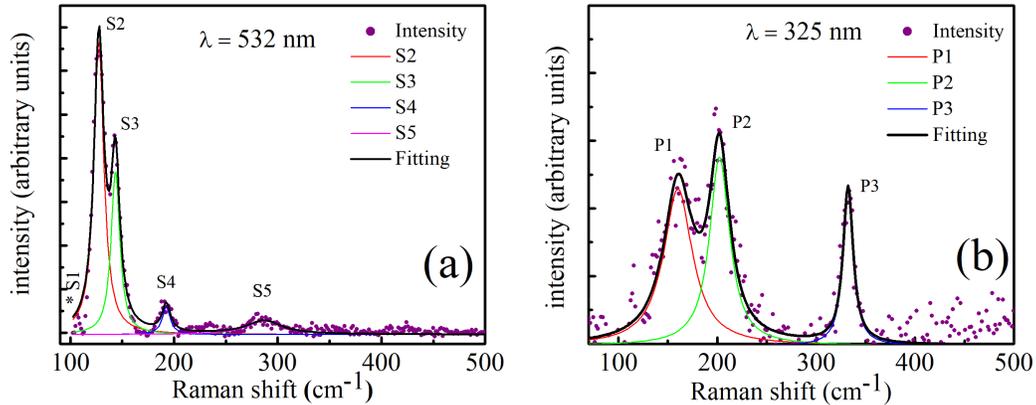

**FIGURE 3:** Raman spectra of $Fe_{0.95}Se_{0.60}Te_{0.40}$ measured at (a) 532 nm and (b) 325 nm excitation wavelengths, along with the peak de-convolution using Lorentzian line width.

The Raman spectra obtained at room temperature with laser wavelength of 532 and 325 nm are shown in Figs. 3(a) and 3(b), respectively. According to the group theory analysis for P4/nmm space group, the Γ point phonon modes of Fe(Se,Te) are given by: $\Gamma = A_{1g} + 2A_u + B_{1g} + 2E_g + 2E_u$. Out of these, there are 4 Raman active phonon modes, namely, $A_{1g}$, $B_{1g}$ and 2 $E_g$ in 100–600 cm$^{-1}$ range [11] while, remaining are IR active modes.

In Fig 3(a), 5 peaks have been obtained where peak maxima have been denoted at 106 (starred, not-fitted), 128 (S2), 143 (S3), 233 (S4), and 285 (S5) cm$^{-1}$ [12, 13]. The chalcogen vibration mode $A_{1g}$ (Te, Se) is visible in the obtained spectra along with Fe vibration mode $B_{1g}$ (Fe) at 128 and 143 cm$^{-1}$, respectively. The $E_g$ (Te, Se) and $E_g$ (Fe) modes are assigned to 106 and 233 cm$^{-1}$, respectively. This assignment is consistent with the reported experimental and calculated phonon frequencies [11, 12]. The mode S5 (285 cm$^{-1}$) has been identified due to hexagonal face of FeSe [12, 14]. The Raman spectrum measured with 325 nm is shown in Fig. 3(b) where only 3 modes at 161 (P1), 201 (P2) and 332 (P3) cm$^{-1}$ are obtained [11, 14]. In this case, $A_{1g}$ (Te, Se) and $B_{1g}$ (Fe) are assigned to P1 and P2 respectively. The $E_g$ (Te,Se) mode is completely absent in this case. The mode arising due to in-plane motion of Fe atoms, i.e., $E_g$ (Fe) has been assigned to peak P3.

## CONCLUSIONS

In conclusion, we have successfully prepared single crystals of $Fe_{0.95}Se_{0.6}Te_{0.4}$ by the modified Bridgman technique. The x-ray diffraction shows presence of only (*0 0 l*) Bragg peaks, which is indicative of good quality of planer crystal growth. The SEM image at 500X magnification shows layered structure of crystal planes. The Raman spectra analysis shows peaks confirming P4/nmm space group formation. The superconducting $T_C$ was determined at 14 K by dc magnetization measurement (ZFC-FC) during heat cycle and further confirmed by the resistivity measurement. For calculating $H_{c1}(0)$ we have analyzed different M-H curves and determined $H_{c1}(T)$ with deviation from the linearity criteria. By extrapolating this data we have determined the value of $H_{c1}(0)$ about 360 Oe and $H_{c1}^* = 1123$ Oe. The temperature exponent of resistivity was found to be ~0.6. This marked a strong deviation from the Fermi liquid behaviour displayed by the metals at low temperature.

## ACKNOWLEDGMENTS

AS and VKM thank the MHRD and SERB-DST, India for the fellowship. The authors acknowledge the Physics Department, Nano Research facility (NRF), and Central Research Facility (CRF), at IIT Delhi for providing research facilities: Glass Blowing, EDX, SEM, XRD, SQUID, PPMS and Raman. We thank Rishabh Shukla, Ajay Kumar and Priyanka Nehla for useful discussions and help during the measurements. RSD gratefully acknowledges the financial support from BRNS through DAE Young Scientist Research Award, project No. 34/20/12/2015/BRNS.

## REFERENCES

1.  Y. Kamihara, T. Watanabe, M. Hirano and H. Hosono, J. Am. Chem. Soc. **130**, 3296-3297 (2008).
2.  F. C. Hsu, J. Y. Luo, K. W. Yeh, T. K. Chen, T. W. Huang, P. M. Wu, Y. C. Lee, Y. L. Huang, Y. Y. Chu, D.C. Yan and M. K. Wu, Proceedings of the National Academy of Sciences **105**, 14262-14264 (2008).
3.  P. Zhang, K. Yaji, T. Hashimoto, Y. Ota, T. Kondo, K. Okazaki, Z. Wang, J. Wen, G. D. Gu, H. Dinga and S. Shin, Science **360,** 182-186 (2018).
4.  B. C. Sales, A. S. Sefat, M. A. McGuire, R. Y. Jin, D. Mandrus and Y. Mozharivskyj, Phys. Rev. B **79,** 094521-094526 (2009).
5.  K. Horigane, N. Takeshita, C. H. Lee, H. Hiraka and K. Yamada, J. Phys. Soc. Jpn. **78**, 063705-063708 (2009).
6.  Y. Mizuguchi, F. Tomioka, S. Tsuda, T. Yamaguchi and Y. Takano, Journal of the Physical Society of Japan **78,** 074712-074717 (2009).
7.  K. W. Yeh, T. W. Huang, Y. L. Huang, T. K. Chen, F. C. Hsu, P. M. Wu, Y. C. Lee, Y. Y. Chu, C. L. Chen, J. Y. Luo and D. C. Yan, Europhysics Letters **84**, 37002-37006 (2008).
8.  P. K. Maheshwari, B. Gahtori, A. Gupta and V. P. S. Awana, AIP Advances **7,** 015006-015015 (2017).
9.  C. S. Yadav and P. L. Paulose, New Journal of Physics. **11**, 103046-103056 (2009).
10. S. Hosoi, K. Matsuura, K. Ishida, H. Wang, Y. Mizukami, T. Watashige, S. Kasahara, Y. Matsuda, and T. Shibauchi, Proc. Natl. Acad. Sci. U.S.A. **113**, 8139-8143 (2016).
11. K. Okazaki, S. Sugai, S. Niitaka, and H. Takagi, Physical Review B **83**, 035103-035112 (2011).
12. R. A. Zargar, A. K. Hafiz and V. P. S. Awana, AIP Conference Proceedings **1675**, 020044-020048 (2015).
13. C. S. Lopes, C. E. Foerster, F. C. Serbena, P. R. Junior, A. R. Jurelo, J. L. Pimentel Junior, P. Pureur and A. L. Chinelatto, Supercond. Sci. Technol. **25,** 025014-025021 (2012).
14. P. Kumar, A. Kumar, S. Saha, D. V. S. Muthu, J. Prakash, S. Patnaik, U. V. Waghmare, A. K. Ganguli, A. K. Sood, Solid State Communications **150**, 557-560 (2010).